\documentclass[aps,prb,reprint,superscriptaddress,nobibnotes]{revtex4-2}

\usepackage{physics}
\usepackage{amssymb}
\usepackage{amsfonts}
\usepackage{mathtools}
\usepackage{hyperref}
\usepackage{tabularx}
\usepackage{graphicx}

\DeclareMathOperator{\diag}{diag}

\begin{document}

\title{\texorpdfstring{$\mathrm{U}(2)$}{U(2)} Chern-Simons-Ginzburg-Landau Theory of Fractional Quantum Hall Hierarchies}

\author{Taegon Lee}
\affiliation{Department of Physics, Korea Advanced Institute of Science and Technology, Daejeon 34141, Republic of Korea}

\author{Gil Young Cho}
\email{gilyoungcho@kaist.ac.kr}
\affiliation{Department of Physics, Korea Advanced Institute of Science and Technology, Daejeon 34141, Republic of Korea}
\affiliation{Center for Artificial Low Dimensional Electronic Systems, Institute for Basic Science, Pohang 37673, Republic of Korea}

\author{Donghae Seo}
\email{donghae98@postech.ac.kr}
\affiliation{Department of Physics, Pohang University of Science and Technology, Pohang 37673, Republic of Korea}

\begin{abstract}
We construct effective $\mathrm{U}(2)$ Chern-Simons-Ginzburg-Landau theories for Abelian and non-Abelian fractional quantum Hall hierarchies for those which had previously been described only through categorical data or trial wavefunctions. Our framework captures both Abelian hierarchy states built on half-filled Pfaffian-type parents and non-Abelian hierarchies emerging from Abelian states. It reproduces all filling fractions obtained from wavefunction and categorical constructions and, moreover, uniquely determines the corresponding topological orders. We also identify an intriguing particle-hole symmetry relating two hierarchy sequences, one built on a trivial insulator and the other on the $\nu=1$ integer quantum Hall state, which respectively generate the Read-Rezayi sequences and their particle-hole conjugates under the same hierarchy construction.
\end{abstract}

\date{\today}

\maketitle

Quantum Hall systems under uniform magnetic fields host a remarkable variety of topologically ordered phases, ranging from the simplest Abelian Laughlin states \cite{laughlin1983anomalous} to more exotic non-Abelian phases such as the Moore-Read Pfaffian \cite{moore1991nonabelions} and the Read-Rezayi states \cite{read1999beyond}. Understanding the structure and organization of these phases remains a central challenge in modern condensed matter physics, despite the long history and depth of the field. One of the most powerful organizing principles is the hierarchy construction \cite{hansson2017quantum}, which relates distinct fractional quantum Hall plateaus through successive condensation of quasiparticles or quasiholes. Originally developed for Abelian states \cite{haldane1983fractional,halperin1984statistics}, this framework has been extended to encompass non-Abelian phases \cite{bonderson2008fractional,levin2009collective,lan2017hierarchy,zhang2025hierarchy,yutushui2025non-abelian}, revealing a rich and intricate structure within the fractional quantum Hall landscape. Recent advances \cite{yutushui2025non-abelian,zhang2025hierarchy} based on wavefunction and categorical approaches have uncovered unexpected hierarchical connections between Abelian and non-Abelian states, yet a unified effective field-theoretic description remains lacking.  

\begin{figure}[t]
    \centering
    \includegraphics[width=0.8\linewidth]{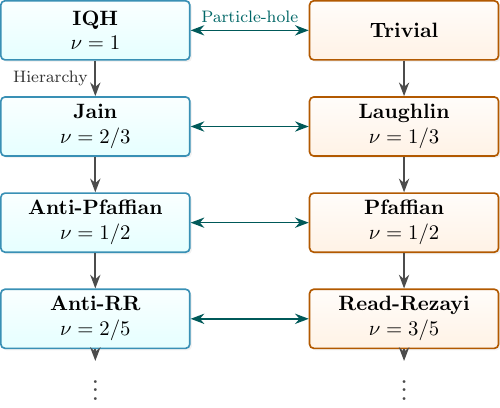}
    \caption{Schematic diagram illustrating the hierarchical and particle-hole relationships between various quantum Hall states. Black and blue arrows indicate hierarchy transitions and particle-hole conjugation, respectively. The left column (cyan boxes) shows the non-Abelian hierarchy sequence emerging from the $\nu = 1$ integer quantum Hall state, while the right column (orange boxes) shows the particle-hole conjugate sequence starting from the trivial insulator.}
    \label{fig:hierarchy_diagram}
\end{figure}

In this Letter, we construct effective $\mathrm{U}(2)$ Chern-Simons-Ginzburg-Landau theories for these newly identified hierarchy sequences, extending beyond the original Abelian hierarchies \cite{wen1992classification} built from Abelian parent states. This framework captures both Abelian hierarchy sequences built on Pfaffian-type parent states (see Table~\ref{tab:hierarchies}) and non-Abelian hierarchy sequences emerging from Abelian parent states (see Fig.~\ref{fig:hierarchy_diagram}). These results are in precise correspondence with hierarchy constructions previously developed in wavefunction and category-theoretic approaches. For example, the Abelian sequence obtained via quasihole condensation from the Pfaffian states in Table~\ref{tab:hierarchies} has been discussed in both wavefunction constructions \cite{levin2009collective} and the stack-and-condense framework \cite{zhang2025hierarchy}. Likewise, the non-Abelian anti-Read-Rezayi hierarchy shown in Fig.~\ref{fig:hierarchy_diagram} has been proposed within the wavefunction approach \cite{yutushui2025non-abelian}.

We find that $\mathrm{U}(2)$ Chern-Simons-Ginzburg-Landau theories and their descendants furnish a unified field-theoretic description of these hierarchy sequences. In this framework, hierarchy states arise from parent states via the development of a finite density of excitations, which subsequently develop their own topological order. When the relevant excitations break the $\mathrm{U}(2)$ gauge symmetry of a non-Abelian parent state down to $\mathrm{U}(1) \times \mathrm{U}(1)$, the resulting daughter states are naturally Abelian, generating Abelian hierarchies from non-Abelian parents. This construction reproduces the Abelian hierarchies associated with the Pfaffian \cite{moore1991nonabelions}, anti-Pfaffian \cite{levin2007particle-hole,lee2007particle-hole}, PH-Pfaffian \cite{son2015is}, bosonic Pfaffian \cite{fradkin1998chern-simons}, and Read-Rezayi states \cite{read1999beyond}. In contrast, when the $\mathrm{U}(2)$ gauge symmetry remains unbroken, the daughter states can remain non-Abelian (depending on the Chern-Simons level), thereby providing a natural route to non-Abelian hierarchies. Within this framework, the resulting topological orders obtained from the $\nu = 1$ integer quantum Hall state are in full agreement with recent wavefunction-based analyses \cite{yutushui2025non-abelian}. Finally, we show that the Read-Rezayi states arise as hierarchical fractional quantum Hall states of a trivial insulating phase within the same framework, and identify a particle-hole symmetric relation with the hierarchies from the integer quantum Hall state.

\begin{table*}[t]
    \centering
    \caption{Summary of Abelian hierarchical fractional quantum Hall states derived from the Chenr-Simons-Ginzburg-Landau framework. The topological order of each state is characterized by $K$ matrix $K_n$ and the charge vector $\mathbf{t}$. Here, $\nu_n$ is the filling fraction, $c_-$ is the chiral central charge, and $D_n$ is the total quantum dimension. For other hierarchical sequences not shown in this table, see the Supplemental Material \cite{supp}.}
    \begin{tabular*}{\textwidth}{@{\extracolsep{\fill}} l|c c c c c @{}}
        \hline \hline
        Parent state & $K_n$ & $\mathbf{t}$ & $\nu_n$ & $c_-$ & $D_n$ \\
        \hline 
        
        \shortstack[l]{Pfaffian \\ (Quasihole)} & 
        $\begin{pmatrix}
        -1 & -3 & -1 \\
        -3 & -1 & 0 \\
        -1 & 0 & -2n
        \end{pmatrix}$ & $\begin{pmatrix} 1\\ 1 \\ 0 \end{pmatrix}$ & $\frac{8 n}{16 n + 1}$ & $0$ & $\sqrt{2 (16 n + 1)}$ \\

        \shortstack[l]{Pfaffian \\ (Quasiparticle)} & 
        $\begin{pmatrix}
        -1 & 0 & 1 & 0 \\
        0 & 2n & -2n+1 & -1 \\
        1 & -2n+1 & 2n & -1 \\
        0 & -1 & -1 & 4
        \end{pmatrix}$
        & $\begin{pmatrix} -1 \\ 0 \\ 0 \\ 0 \end{pmatrix}$ & $\frac{8 n - 1}{16 n - 3}$ & $3$ & $\sqrt{2 (16 n - 3)}$ \\
        
        \shortstack[l]{Anti-Pfaffian \\ (Quasihole)} & 
        $\begin{pmatrix}
        1 & 3 & -1 \\
        3 & 1 & 0 \\
        -1 & 0 & 2n
        \end{pmatrix}$
        & $\begin{pmatrix} 1 \\ 1 \\ 0 \end{pmatrix}$ & $\frac{8 n + 1}{16 n + 1}$ & $1$ & $\sqrt{2 (16 n + 1)}$ \\

        \shortstack[l]{Anti-Pfaffian \\ (Quasiparticle)} & 
        $\begin{pmatrix}
        1 & 0 & -1 & 0 \\
        0 & -2n & 2n-1 & -1 \\
        -1 & 2n-1 & -2n & -1 \\
        0 & -1 & -1 & -4
        \end{pmatrix}$
        & $\begin{pmatrix} 1 \\ 0 \\ 0 \\ 0 \end{pmatrix}$ & $\frac{8 n - 2}{16 n - 3}$ & $-2$ & $\sqrt{2 (16 n - 3)}$ \\

        \shortstack[l]{PH-Pfaffian \\ (Quasihole)} & 
        $\begin{pmatrix}
        0 & 2 & -1 & -1 \\ 
        2 & 0 & -1 & 0 \\
        -1 & -1 & 3 & 0 \\
        -1 & 0 & 0 & -2n 
        \end{pmatrix}$
        & $\begin{pmatrix} 0 \\ 0 \\ -1 \\ 0 \end{pmatrix}$ & $\frac{8n}{16n+1}$ & $0$ & $\sqrt{2(16n+1)}$ \\

        \shortstack[l]{PH-Pfaffian \\ (Quasiparticle)} & 
        $\begin{pmatrix}
        0 & 2 & -1 & -1 \\ 
        2 & 0 & -1 & 0 \\
        -1 & -1 & 3 & 0 \\
        -1 & 0 & 0 & 2n 
        \end{pmatrix}$
        & $\begin{pmatrix} 0 \\ 0 \\ -1 \\ 0 \end{pmatrix}$ & $\frac{8 n}{16 n - 1}$ & $2$ & $\sqrt{2 (16 n - 1)}$ \\
        \hline \hline
    \end{tabular*}
    \label{tab:hierarchies}
\end{table*}

\textbf{1.~Abelian hierarchies of Pfaffian state.} We begin with the Abelian daughter states of the Pfaffian phase, for example at filling fractions $\nu = 8/17$ and $7/13$, obtained via condensation of minimally charged non-Abelian anyons into incompressible liquids~\cite{levin2009collective}.

The Pfaffian state is described by the $\mathrm{U}(2)_{2, -8} \times \mathrm{U}(1)_1$ Chern-Simons theory \cite{fradkin1998chern-simons}. Its anyon contents are labeled by $(j, n)$, where $2j$ and $n$ are integers satisfying $j + n / 2 \in \mathbb{Z}$. The indices $j$ and $n$ respectively denote the spin-$j$ representation of $\mathrm{SU}(2)$ and the charge-$n$ representation of $\mathrm{U}(1)$. The charge of the anyon $(j, n)$ is given by $e n / 4$ where $e$ is the electron charge. Fusing two minimally charged $(1 / 2, 1)$ anyons yield $(1 / 2, 1) \otimes (1 / 2, 1) = (0, 2) \oplus (1, 2)$. Since the former fusion channel gives an anyon carrying a spin singlet while the latter a triplet, each fusion channel is referred to as ``paramagnetic'' and ``ferromagnetic,'' respectively \cite{shi2025dopinglattice}. We will adopt the terminologies in the following.

To describe the hierarchy transition from the Pfaffian state, we introduce anyons as a scalar field in the appropriate representation~\cite{zhang1989effective,read1989order,lee1991collective,lopez1991fractional,goldman2019landau,shi2025dopinglattice}. The $(1/2,1)$ anyon is thus represented by a scalar field $\Phi$ in the fundamental representation of $\mathrm{U}(2)$~\cite{shi2025dopinglattice}. The Lagrangian is given by
\begin{align} \label{eq:pfaffian_csgl}
    \mathcal{L} &= - \frac{2}{4 \pi} \Tr\!\left[a d a + \frac{2}{3} a^3\right] + \frac{3}{4 \pi} (\Tr a) d (\Tr a) \nonumber \\
    &\quad + \frac{1}{2 \pi} A d (\Tr a) + \mathrm{CS}[A, g] + \mathcal{L}[\Phi, a],
\end{align}
where $a$ is a dynamical $\mathrm{U}(2)$ gauge field, $A$ is the background electromagnetic field, and $\mathrm{CS}[A, g] = \frac{1}{4 \pi} A d A + 2 \mathrm{CS}_g$. Here, $\mathrm{CS}_g$ is the gravitational Chern-Simons term with the chiral central charge $c_- = \frac{1}{2}$ \cite{seiberg2016duality}. The matter sector is given by $\mathcal{L}[\Phi, a] = \Tr[(D\Phi)^\dagger (D\Phi)] - V(\Phi)$, where $D\Phi = \partial \Phi - i a \Phi$ and $V(\Phi)$ is an appropriate potential. 

To obtain the Abelian hierarchy, we assume that the ferromagnetic fusion channel of the $(1/2,1)$ anyon is favored, leading to spontaneous breaking of the $\mathrm{U}(2)$ gauge symmetry down to $\mathrm{U}(1)\times \mathrm{U}(1)$~\cite{shi2025dopinglattice}. Without loss of generality, we take the spin polarization to align along the $z$-direction. In the low-energy regime, the $\mathrm{U}(2)$ gauge field is then reduced to $a \to \mathrm{diag}(a_\uparrow, a_\downarrow)$, where $a_\uparrow$ and $a_\downarrow$ are $2\pi$-quantized $\mathrm{U}(1)$ gauge fields. The Lagrangian in Eq.~\eqref{eq:pfaffian_csgl} thus reduces to
\begin{align} \label{eq:pfaffian_quasihole_before_dual}
    \mathcal{L} &= \frac{1}{4 \pi} (a_\uparrow d a_\uparrow + a_\downarrow d a_\downarrow) + \frac{3}{2 \pi} a_\uparrow d a_\downarrow + \frac{1}{2 \pi} A d (a_\uparrow + a_\downarrow) \nonumber \\
    & \quad+ \mathrm{CS}[A, g] + \mathcal{L}[\Phi_\uparrow, a_\uparrow],
\end{align}
where $\Phi = (\Phi_\uparrow, \Phi_\downarrow)^\mathsf{T}$. Here, we have assumed that $\Phi_\downarrow$ has zero density and integrated it out. Since Eq.~\eqref{eq:pfaffian_quasihole_before_dual} describes an Abelian Chern--Simons theory, we can straightforwardly apply the standard hierarchy construction for Abelian sequences \cite{wen1992classification}, as detailed below.

We begin by constructing the quasihole hierarchy of the Pfaffian state. To this end, we perform a particle-vortex duality transformation on the current of $\Phi_\uparrow$ by introducing a dynamical $\mathrm{U}(1)$ gauge field $\alpha$. We then supplement a level-$2n$ self Chern-Simons term of $\alpha$, yielding
\begin{align} \label{eq:pfaffian_quasihole}
    \mathcal{L} &= \frac{1}{4 \pi} (a_\uparrow d a_\uparrow + a_\downarrow d a_\downarrow) + \frac{3}{2 \pi} a_\uparrow d a_\downarrow + \frac{1}{2 \pi} A d (a_\uparrow + a_\downarrow) \nonumber \\
    &\quad  + \mathrm{CS}[A, g] + \frac{1}{2 \pi} a_\uparrow d \alpha + \frac{2 n}{4 \pi} \alpha d \alpha.
\end{align}
In the $K$-matrix formulation of Abelian topological orders \cite{wen1992classification}, the Lagrangian in Eq.~\eqref{eq:pfaffian_quasihole} is represented by
\begin{align} \label{eq:pfaffian_quasihole_kmatrix}
    K_n = 
    \begin{pmatrix}
        -1 & -3 & -1 \\
        -3 & -1 & 0 \\
        -1 & 0 & -2n
   \end{pmatrix}, \quad 
    \mathbf{t} = 
    \begin{pmatrix}
        1 \\ 1 \\ 0
   \end{pmatrix},
\end{align}
where $\mathbf{t}$ is the charge vector. From Eq.~\eqref{eq:pfaffian_quasihole_kmatrix}, the filling fraction, the chiral central charge, and the total quantum dimension are computed as $\nu_n = \frac{8 n}{16 n + 1}$, $c_{-} = 0$, and $D_n = \sqrt{2 (16 n + 1)}$, respectively. Here, $\mathrm{CS}[A,g]$ term contributes to $c_{-}$ by the factor of $+1$. Moreover, the resulting topological order is generated by the abelian anyon $ (0,1,2n+1)^T$, which has topological spin $\frac{16 n - 1}{2 (16 n + 1)}$ and electric charge $\frac{1}{16 n + 1}$ in units of the electron charge. This agrees with results from category-theoretic approaches~\cite{zhang2025hierarchy}.

Notably, we can manifestly identify the resulting Abelian hierarchical states as those built on the strong-pairing phase at filling fraction $\nu = 1/2$, originally proposed based on the quasihole statistics in the wavefunction construction~\cite{levin2009collective}. Under an $\mathrm{SL}(3, \mathbb{Z})$ transformation~\cite{wen1992classification}, Eq.~\eqref{eq:pfaffian_quasihole_kmatrix} becomes 
\begin{align}
    K_n \to 
    \begin{pmatrix}
        -1 & 0 & 1 \\
        0 & 8 & -3 \\
        1 & -3 & -2n
    \end{pmatrix}, \quad 
    \mathbf{t} \to 
    \begin{pmatrix}
        -1 \\ 2 \\ 0
    \end{pmatrix}. \nonumber
\end{align}
This can be interpreted as follows. The upper $2\times 2$ block of the above $K_n$, namely $\mathrm{diag}(-1,8)$, together with $\mathrm{CS}[A,g]$ in Eq.~\eqref{eq:pfaffian_quasihole}, describes the strong-pairing phase at $\nu = 1/2$. From this state, the Abelian anyon $\mathbf{l} = (-1,3)^\mathsf{T}$ with charge $-e/4$ condenses into an incompressible quantum Hall state at $\nu = 2n$, producing the resulting $K$ matrix and charge vector $\mathbf{t}$. 

Within the same framework, we construct the quasiparticle hierarchy. To this end, we first dualize the current of $\Phi_\uparrow$ in Eq.~\eqref{eq:pfaffian_quasihole_before_dual}, as before. We then introduce a level-$(-2n)$ self Chern-Simons term for $\alpha$. To reproduce the topological data reported in \cite{levin2009collective,zhang2025hierarchy}, we further introduce a current $\star j' = \frac{1}{2\pi} d\beta$ that carries unit charge under $\alpha$, along with a level-$(-4)$ Chern-Simons term for $\beta$, where $\beta$ is an emergent $\mathrm{U}(1)$ gauge field. In the $K$-matrix formalism, the resulting Lagrangian is given by
\begin{equation}\label{eq:pfaffian_quasiparticle_kmatrix1}
    K_n = 
    \begin{pmatrix}
        -1 & -3 & -1 & 0 \\
        -3 & -1 & 0 & 0 \\ 
        -1 & 0 & 2n & -1 \\
        0 & 0 & -1 & 4 
    \end{pmatrix},
    \quad
    \mathbf{t} = 
    \begin{pmatrix}
        1 \\ 1 \\ 0 \\ 0
    \end{pmatrix}.
\end{equation}
In the resulting theory, there exists a charge-neutral boson of order 2 labeled by $\mathbf{l}_n = (0,1,-2n+1,-1)^\mathsf{T}$, which should condense. It carries topological spin $s_{\mathbf{l}_n} = n$ and electric charge $q_{\mathbf{l}_n} = 0$, but cannot condense directly since its topological spin is not trivial. To circumvent this, we stack an appropriate number of $\nu=\pm 1$ trivial blocks onto the $K$-matrix and condense a composite of $\mathbf{l}_n$ with fermions from these trivial sectors. Details are provided in the Supplemental Material~\cite{supp}. After the condensation followed by an $\mathrm{SL}(N,\mathbb{Z})$ transformation, Eq.~\eqref{eq:pfaffian_quasiparticle_kmatrix1} becomes
\begin{equation}\label{eq:pfaffian_quasiparticle_kmatrix2}
    K_{n} = 
    \begin{pmatrix}
        -1 & 0 & 1 & 0 \\
        0 & 2n & 1-2n & -1 \\
        1 & 1-2n & 2n & -1 \\
        0 & -1 & -1 & 4
    \end{pmatrix},
    \quad
    \mathbf{t} =
    \begin{pmatrix}
        -1 \\ 0 \\ 0 \\ 0
    \end{pmatrix}. \nonumber
\end{equation}
The filling fraction, the chiral central charge, and the total quantum dimension are computed as $\nu_n = \frac{8n - 1}{16 n - 3}$, $c_{-} = 3$, and $D_n = \sqrt{2 (16 n - 3)}$. The resulting phase has an Abelian anyon theory generated by $(0,0,-1,2)^\mathsf{T}$ which has topological spin $\frac{16 n - 1}{2 (16 n - 3)}$ and electric charge $\frac{e}{16 n - 3}$. Again, this agrees with \cite{levin2009collective, zhang2025hierarchy}.

Our framework can be readily extended to other non-Abelian fractional quantum Hall states, including the anti-Pfaffian, PH-Pfaffian, and bosonic Pfaffian states. As the analysis closely parallels that of the Pfaffian hierarchies, we defer the details to the Supplemental Material~\cite{supp} and summarize the results in Table~\ref{tab:hierarchies}. Remarkably, our results are in precise agreement with a recent category-theoretic approach~\cite{zhang2025hierarchy}, thereby providing a field-theoretic description of the Abelian hierarchy sequences of Pfaffian-type states. Furthermore, this framework enables systematic generalizations that yield new Abelian hierarchy states. For example, in the quasiparticle hierarchy of the Pfaffian state, the level-$(-4)$ Chern-Simons term for $\beta$ can be generalized to a level-($-2m$) term, thereby yielding a new sequence of Abelian hierarchy states at filling factor $\nu_{n,m} = \frac{8 m n - 2}{16 m n - m - 4}$.

\textbf{2.~Non-Abelian hierarchies of Jain state.} In a recent wavefunction analysis~\cite{yutushui2025non-abelian}, it was proposed that condensation of Abelian anyons in an Abelian state can give rise to non-Abelian hierarchy states. More specifically, successive condensations of Laughlin quasiholes---anyon excitations associated with flux insertion---generate a sequence of $\mathbb{Z}_k$ parafermion anti-Read-Rezayi states starting from the $\nu = 1$ integer quantum Hall state (see Fig.~\ref{fig:hierarchy_diagram}). Since the transition from the $\nu = 1$ integer quantum Hall state to the $\nu = 2/3$ Jain state is well described within the standard Abelian hierarchy framework \cite{wen1992classification}, we focus on a field-theoretic description that captures the non-Abelian hierarchy sequences emerging from the $\nu = 2/3$ Jain state. 

To begin, we propose that the $\nu = 2/3$ Jain state admits an $\mathrm{U}(2)_{-1,6}$ Chern-Simons theory description. Since the $\mathrm{SU}(2)$ Chern-Simons level is $-1$, the theory is effectively Abelian. The anyon content is generated by the excitation $(j,n)=(1/2,1)$, which carries topological spin $-1/6$ and electric charge $e/3$, producing the full set of allowed anyons $(0,0)$, $(1/2,1)$, $(0,2)$, $(1/2,3)$, $(0,4)$, and $(1/2,5)$ in the theory. Further, the chiral central charge is $c_- = 0$. Since a fermionic topological order is fully characterized by its anyon content and chiral central charge~\cite{lan2016theory,bruillard2017fermionic,cho2023classification}, this establishes that the $\mathrm{U}(2)_{-1,6}$ Chern-Simons theory indeed describes the Jain state.

In the $\mathrm{U}(2)_{-1,6}$ Chern-Simons description, the Laughlin quasihole corresponds to the $(j,n)=(0,2)$ anyon. Let $\phi$ denote the scalar field sourcing this quasihole, with gauge-covariant derivative $D\phi = \partial \phi - i(\Tr a) \phi$. Since $\phi$ couples only to the Abelian sector, its direct condensation cannot generate a non-Abelian Chern-Simons term. To overcome this limitation, we instead consider the proliferation of minimally charged anyons $(j,n)=(1/2,1)$, which preferentially fuse into the paramagnetic channel, i.e., the $(0,2)$ sector. The resulting theory is then 
\begin{align}
    \mathcal{L} &= \frac{1}{4 \pi} \Tr\!\left[a d a + \frac{2}{3} a^3\right] - \frac{2}{4 \pi} (\Tr a) d (\Tr a) \nonumber \\
    &\quad + \frac{1}{2 \pi} A d (\Tr a) + \mathcal{L}[\Phi, a]. \nonumber
\end{align}
To construct the hierarchical state, we attach flux to $\Phi$ by modifying its gauge-covariant derivative as $D\Phi = \partial \Phi - i a \Phi + i \Phi b$, where $b$ is an additional $\mathrm{U}(2)$ gauge field, and by introducing a Chern-Simons term for $b$. The simplest bosonic $\mathrm{U}(2)$ Chern-Simons term with vanishing chiral central charge is given by $\frac{1}{4\pi}\Tr\!\left[b d b + \frac{2}{3} b^3\right] - \frac{1}{4\pi}(\Tr b) d(\Tr b)$. The full theory then becomes 
\begin{align} \label{eq:csgl_jain}
    \mathcal{L} &= \frac{1}{4 \pi} \Tr\!\left[a d a + \frac{2}{3} a^3\right] - \frac{2}{4 \pi} (\Tr a) d (\Tr a) \nonumber \\
    &\quad + \frac{1}{2 \pi} A d (\Tr a) + \Tr\!\left[|\partial \Phi - i a \Phi + i \Phi b|^2\right] - V(\Phi) \nonumber \\
    &\quad + \frac{1}{4 \pi} \Tr\!\left[b d b + \frac{2}{3} b^3\right] - \frac{1}{4 \pi} (\Tr b) d (\Tr b). 
\end{align}
Notably, when $\langle \Phi \rangle = 0$, the $\mathrm{U}(2)$ Chern-Simons term for $b$ describes a trivial topological order, and Eq.~\eqref{eq:csgl_jain} reduces to the $\nu = 2/3$ Jain state. Upon condensation of $\Phi$, the theory undergoes a transition to a different topological order, such that the condensation transition realizes the quantum Hall hierarchy transition. The condensation $\langle \Phi \rangle \neq 0$ triggers a Higgs mechanism in which the gauge field combination $a - b$ becomes massive, imposing the constraint $a = b$ in the low-energy regime. The Lagrangian then reduces to
\begin{align} \label{eq:anti-pfaffian}
    \mathcal{L} &= \frac{2}{4 \pi} \Tr\!\left[a d a + \frac{2}{3} a^3\right] - \frac{3}{4 \pi} (\Tr a) d (\Tr a) \nonumber \\
    &\quad + \frac{1}{2 \pi} A d (\Tr a), \nonumber
\end{align}
which describes the anti-Pfaffian state. This establishes the anti-Pfaffian state as a hierarchical fractional quantum Hall state emerging from the $\nu = 2/3$ Jain state.

By iterating the above procedure, we reproduce the full sequence of non-Abelian hierarchies constructed in the wavefunction approach~\cite{yutushui2025non-abelian}. The $\mathbb{Z}_k$ anti-Read-Rezayi state is described by
\begin{align}
    \mathcal{L} &= \frac{k}{4 \pi} \Tr\!\left[a d a + \frac{2}{3} a^3\right] - \frac{k + 1}{4 \pi} (\Tr a) d (\Tr a) \nonumber \\
    &\quad + \frac{1}{2 \pi} A d (\Tr a) + \mathcal{L}[\Phi, a]. \nonumber
\end{align}
Attaching the minimal $\mathrm{U}(2)$ Chern-Simons flux to $\Phi$ by introducing $\frac{1}{4 \pi} \Tr\!\left[b d b + \frac{2}{3} b^3\right] - \frac{1}{4 \pi} (\Tr b) d (\Tr b)$, and subsequently condensing $\Phi$, yields
\begin{align}
    \mathcal{L} &= \frac{k + 1}{4 \pi} \Tr\!\left[a d a + \frac{2}{3} a^3\right] - \frac{k + 2}{4 \pi} (\Tr a) d (\Tr a) \nonumber \\
    &\quad + \frac{1}{2 \pi} A d (\Tr a), \nonumber
\end{align}
which describes the $\mathbb{Z}_{k+1}$ anti-Read-Rezayi state. A similar construction applies to the non-Abelian hierarchy sequence built on the $\nu = 1/3$ Laughlin state~\cite{yutushui2025non-abelian}, which we defer the details to the Supplemental Material~\cite{supp}.

We note that there exist similar approaches to constructing non-Abelian topological phases based on Chern-Simons dualities~\cite{goldman2019landau,goldman2020non-abelian}. In \cite{goldman2019landau}, the bosonic $\mathbb{Z}_k$ Read-Rezayi state is obtained from $k$ layers of $\nu = 1/2$ Laughlin states by mapping the $\mathrm{U}(1)_2$ theory to $\mathrm{SU}(2)_1$ and then subsequently Higgsing the $\mathrm{SU}(2)_1 \times \mathrm{SU}(2)_1 \times \cdots$ theory to the $\mathrm{SU}(2)_k$ theory. In \cite{goldman2020non-abelian}, a $\mathrm{U}(k)_2$ theory is similarly derived via a dual fermionic description of the $\nu = 1/2$ Laughlin state and exciton pairing. In our case, a trivial bosonic layer is introduced so that the transition does not involve nontrivial intermediate phases.

\textbf{3.~Read-Rezayi hierarchies.} Motivated by the hierarchy construction of the anti-Read-Rezayi states, we now formulate a Chern-Simons-Ginzburg-Landau theory for the Read-Rezayi states emerging from a trivial insulating parent at $\nu = 0$. The trivial insulator can be viewed as the particle-hole conjugate of the $\nu = 1$ integer quantum Hall state. It is therefore natural to expect that applying an analogous construction yields the Read-Rezayi sequence, which is the particle-hole conjugate of the anti-Read-Rezayi sequence.

The trivial insulator is described by: 
\begin{align}
    \mathcal{L} &= \frac{1}{4 \pi} (\Tr a) d (\Tr a) + \frac{1}{2 \pi} A d (\Tr a) + \mathrm{CS}[A, g] + \mathcal{L}[\Phi, a]. \nonumber
\end{align}
We then introduce $- \frac{1}{4\pi}\Tr\!\left[b d b + \frac{2}{3} b^3\right] + \frac{1}{4\pi}(\Tr b) d(\Tr b)$ as before \footnote{Note that the overall sign is reversed compared to the previous case; nevertheless, the Chern-Simons term still describes a trivial bosonic topological order.}, with $\Phi$ transforming in the bifundamental representation of $a$ and $b$, and subsequently condense $\Phi$. Repeating this procedure $k$ times, we obtain   
\begin{align}
    \mathcal{L} &= - \frac{k}{4 \pi} \Tr\!\left[a d a + \frac{2}{3} a^3\right] + \frac{k + 1}{4 \pi} (\Tr a) d (\Tr a) \nonumber \\
    &\quad + \frac{1}{2 \pi} A d (\Tr a) + \mathrm{CS}[A, g], \nonumber
\end{align}
which is precisely the effective theory of the $\mathbb{Z}_k$ Read-Rezayi state. All the hierarchy states from the trivial insulator are the particle-hole conjugates of the hierarchical anti-Read-Rezayi states discussed previously. The schematic relations are illustrated in Fig.~\ref{fig:hierarchy_diagram}.

\textbf{4.~Conclusion.} We have presented a systematic construction of $\mathrm{U}(2)$ Chern-Simons-Ginzburg-Landau theories for hierarchical quantum Hall states, elucidating the connection between Abelian and non-Abelian topological orders. Within a unified field-theoretic framework, our approach captures both Abelian hierarchies built on Pfaffian-type non-Abelian parent states, and non-Abelian hierarchies emerging from Abelian parents. The resulting filling fractions and topological orders are in precise agreement with previous wavefunction~\cite{levin2009collective,yutushui2025non-abelian} and category-theoretic studies~\cite{zhang2025hierarchy}. We also discovered particle-hole symmetry between non-Abelian hierarchy sequences emerging from the trivial topological order and the $\nu = 1$ integer quantum Hall state.

Our $\mathrm{U}(2)$ Chern-Simons-Ginzburg-Landau description of hierarchy states closely parallels that of anyon superconductivity in non-Abelian fractional anomalous Hall systems~\cite{shi2025dopingfractional,pichler2025microscopic,shi2025dopinglattice,zhang2025charge,kuhlenkamp2025robust,wang2025chiral,zhang2025holon,ahn2026superconductivity,seo2026unified}. In both cases, the phases emerge from a parent topological order via condensation of anyonic excitations into incompressible states and share closely related mathematical structures, as emphasized in a recent category-theoretic analysis~\cite{seo2026unified}. The present field-theoretic framework accommodates a broad range of many-body states emerging from condensing anyons, suggesting a rich landscape of anyon-driven phases that warrants systematic exploration in future work.

\begin{acknowledgments}
    This work is financially supported by Samsung Science and Technology Foundation under Project Number SSTF-BA2401-03, the NRF of Korea (Grants No. RS-2026-25479545, RS-2024-00410027, RS-2023-NR119931, RS-2024-00444725, RS-2023-00256050, IRS-2025-25453111, RS-2025-08542968) funded by the Korean Government (MSIT), the Air Force Office of Scientific Research under Award No. FA23862514026, and Institute of Basic Science under project code IBS-R014-D1. T.~L. is partially supported by KAIST Undergraduate Research Program (URP).
\end{acknowledgments}

\bibliography{ref}

\clearpage
\onecolumngrid

\begin{center}
    \large{\bf Supplemental Material for ``\texorpdfstring{$\mathrm{U}(2)$}{U(2)} Chern-Simons-Ginzburg-Landau Theory of Fractional Quantum Hall Hierarchies''}
\end{center}
\vspace{1ex}

\twocolumngrid

\tableofcontents

\section{\texorpdfstring{$\mathrm{SL}(N,\mathbb{Z})$}{\mathrm{SL}(N,Z)} transformation}

\begin{table*}[t]
    \centering
    \caption{Summary of Abelian hierarchical fractional quantum Hall states derived from the Chern-Simons-Ginzburg-Landau framework. The topological order of each state is characterized by a $K$ matrix $K_n$ and a charge vector $\mathbf{t}$. Here, $\nu_n$ is the filling fraction, $c_{-}$ is the chiral central charge, and $D_n$ is the total quantum dimension of the hierarchy states.}
    \begin{tabular*}{\textwidth}{@{\extracolsep{\fill}} l|c c c c c @{}}
        \hline
        Parent state & $K_n$ & $\mathbf{t}$ & $\nu_n$ & $c_-$ & $D_n$ \\
        \hline 
        \shortstack[l]{Pfaffian \\ (Quasihole)} & $\begin{pmatrix} 8 & -1 \\ -1 & -2n \end{pmatrix}$ & $\begin{pmatrix} 2 \\ 0 \end{pmatrix}$ & $\frac{8 n}{16 n + 1}$ & $0$ & $\sqrt{2 (16 n + 1)}$ \\

        \shortstack[l]{Pfaffian \\ (Quasiparticle)} & 
        $\begin{pmatrix}
        -1 & 0 & 1 & 0 \\
        0 & 2n & -2n+1 & -1 \\
        1 & -2n+1 & 2n & -1 \\
        0 & -1 & -1 & 4
        \end{pmatrix}$
        & $\begin{pmatrix} -1 \\ 0 \\ 0 \\ 0 \end{pmatrix}$ & $\frac{8 n - 1}{16 n - 3}$ & $3$ & $\sqrt{2 (16 n - 3)}$ \\
        
        \shortstack[l]{Pfaffian \\ (Quasihole)} & 
        $\begin{pmatrix}
        -1 & 0 & 1 & 0 \\
        0 & -2n & 2n & -1 \\
        1 & 2n & -2n & -1 \\
        0 & -1 & -1 & -4
        \end{pmatrix}$
        & $\begin{pmatrix} 1 \\ 0 \\ 0 \\ 0 \end{pmatrix}$ & $\frac{8n-1}{16n-1}$ & $-1$ & $\sqrt{2(16n-1)}$ \\

        \shortstack[l]{Pfaffian \\ (Quasiparticle)} & 
        $\begin{pmatrix}
        2n & 2n-1 & 0 & 1 & 2n-1 \\
        2n-1 & 2n & 0 & 1 & 2n-1 \\
        0 & 0 & -1 & -3 & 1 \\
        1 & 1 & -3 & -1 & 0 \\
        2n-1 & 2n-1 & 1 & 0 & 2n
        \end{pmatrix}$
        & $\begin{pmatrix} 0 \\ 0 \\ -1 \\ -1 \\ 0 \end{pmatrix}$ & $\frac{8 n - 2}{16 n - 5}$ & $4$ & $\sqrt{2 (16 n - 5)}$ \\
        
        \shortstack[l]{Anti-Pfaffian \\ (Quasihole)} & 
        $\begin{pmatrix}
        1 & 3 & -1 \\
        3 & 1 & 0 \\
        -1 & 0 & 2n
        \end{pmatrix}$
        & $\begin{pmatrix} 1 \\ 1 \\ 0 \end{pmatrix}$ & $\frac{8n+1}{16n+1}$ & $1$ & $\sqrt{2 (16n+1)}$ \\

        \shortstack[l]{Anti-Pfaffian \\ (Quasiparticle)} & 
        $\begin{pmatrix}
        1 & 0 & -1 & 0 \\
        0 & -2n & 2n-1 & -1 \\
        -1 & 2n-1 & -2n & -1 \\
        0 & -1 & -1 & -4
        \end{pmatrix}$
        & $\begin{pmatrix} 1 \\ 0 \\ 0 \\ 0 \end{pmatrix}$ & $\frac{8 n - 2}{16 n - 3}$ & $- 2$ & $\sqrt{2 (16 n - 3)}$ \\

        \shortstack[l]{Bosonic Pfaffian \\ (Quasihole)} & 
        $\begin{pmatrix}
        0 & -2 & -1 & -1 \\ 
        -2 & 0 & -1 & 0 \\
        -1 & -1 & 0 & 0 \\
        -1 & 0 & 0 & -2n
        \end{pmatrix}$
        & $\begin{pmatrix} 0 \\ 0 \\ 1 \\ 0 \end{pmatrix}$ & $\frac{8 n}{8 n + 1}$ & $0$ & $\sqrt{8 n + 1}$ \\

        \shortstack[l]{Bosonic Pfaffian \\ (Quasiparticle)} & 
        $\begin{pmatrix}
        0 & 0 & 1 & 1 & 0 \\
        0 & 2n & 1 & 2n & 1 \\
        1 & 1 & 0 & 0 & 0 \\
        1 & 2n & 0 & 2n & -1 \\
        0 & 1 & 0 & -1 & 4
        \end{pmatrix}$
        & $\begin{pmatrix} 0 \\ 0 \\ 1 \\ 0 \\ 0 \end{pmatrix}$ & $\frac{8 n - 1}{8 n - 2}$ & $3$ & $\sqrt{8 n - 2}$ \\

        \shortstack[l]{PH-Pfaffian \\ (Quasihole)} & 
        $\begin{pmatrix}
        0 & 2 & -1 & -1 \\ 
        2 & 0 & -1 & 0 \\
        -1 & -1 & 3 & 0 \\
        -1 & 0 & 0 & -2n 
        \end{pmatrix}$
        & $\begin{pmatrix} 0 \\ 0 \\ -1 \\ 0 \end{pmatrix}$ & $\frac{8n}{16n+1}$ & $0$ & $\sqrt{2(16n+1)}$ \\

        \shortstack[l]{PH-Pfaffian \\ (Quasiparticle)} & 
        $\begin{pmatrix}
        0 & 2 & -1 & -1 \\ 
        2 & 0 & -1 & 0 \\
        -1 & -1 & 3 & 0 \\
        -1 & 0 & 0 & 2n 
        \end{pmatrix}$
        & $\begin{pmatrix} 0 \\ 0 \\ -1 \\ 0 \end{pmatrix}$ & $\frac{8 n}{16 n - 1}$ & $2$ & $\sqrt{2 ( 16 n - 1)}$ \\

        \shortstack[l]{PH-Pfaffian \\ (Quasihole)} & 
        $\begin{pmatrix}
        0 & 0 & -1 & -1 & 0 \\
        0 & -2n & -1 & -2n & 1 \\
        -1 & -1 & 3 & 0 & 0 \\
        -1 & -2n & 0 & -2n & -1 \\
        0 & 1 & 0 & -1 & -4
        \end{pmatrix}$
        & $\begin{pmatrix} 0 \\ 0 \\ -1 \\ 0 \\ 0 \end{pmatrix}$ & $\frac{8n-1}{16n-1}$ & $-1$ & $\sqrt{2(16n-1)}$ \\

        \shortstack[l]{PH-Pfaffian \\ (Quasiparticle)} & 
        $\begin{pmatrix}
        0 & 0 & -1 & -1 & 0 \\
        0 & 2n-2 & -1 & 2n-1 & -3 \\
        -1 & -1 & 3 & 0 & 0 \\
        -1 & 2n-1 & 0 & 2n & -1 \\
        0 & -3 & 0 & -1 & -4
        \end{pmatrix}$
        & $\begin{pmatrix} 0 \\ 0 \\ 1 \\ 0 \\ 0 \end{pmatrix}$ & $\frac{8 n + 1}{16 n + 1}$ & $1$ & $\sqrt{2 (16 n + 1)}$ \\
        \hline \hline
    \end{tabular*}
\end{table*}

In the main text, to construct the quasiparticle hierarchy of the Pfaffian state, we have condensed a charge-neutral boson of order-$2$ and then performed an $\mathrm{SL}(N,\mathbb{Z})$ transformation. In this section, we explicitly show how this boson can be condensed and determine a suitable $\mathrm{SL}(N,\mathbb{Z})$ transformation matrix. 

The intermediate $K$ matrix and the charge vector for the quasiparticle hierarchy of the Pfaffian state are given by
\begin{equation}\label{eq:pfaffian_quasiparticle_kmatrix1*}
    K_n = 
    \begin{pmatrix}
        -1 & -3 & -1 & 0 \\
        -3 & -1 & 0 & 0 \\ 
        -1 & 0 & 2n & -1 \\
        0 & 0 & -1 & 4 
    \end{pmatrix},
    \quad
    \mathbf{t} = 
    \begin{pmatrix}
        1 \\ 1 \\ 0 \\ 0
    \end{pmatrix}.
\end{equation}
The anyon to be condensed is labeled by an integer vector $\mathbf{l}_n = (0,1,-2n+1,-1)^\mathsf{T}$. It carries zero electric charge, $q_n = \mathbf{t}^\mathsf{T} K_n^{-1} \mathbf{l}_n =0$, and has order $2$. Its topological spin is $s_{\mathbf{l}_n} = \frac{1}{2} \mathbf{l}_n^\mathsf{T} K_n^{-1} \mathbf{l}_n =n \in \mathbb{Z}$, and it is therefore a condensable boson. However, its topological spin is not trivial, which obstructs a direct condensation, as noted in Ref.~\cite{zhang2025hierarchy}. This obstruction arises from a residual Chern-Simons term for $\beta$ that remains after integrating out the dynamical gauge fields.

We now generalize the construction in Ref.~\cite{zhang2025hierarchy} to make the topological spin of the condensate exactly zero and to construct a suitable $\mathrm{SL}(N,\mathbb{Z})$ transformation matrix. First, we stack $2s_{\mathbf{l}_n}=2n$ copies of $\nu=+1$ and $2s_{\mathbf{l}_n}=2n$ copies of $\nu=-1$ blocks onto the $K$ matrix, which do not alter the topological order. Then the $K$ matrix and the charge vector given in Eq.~\eqref{eq:pfaffian_quasiparticle_kmatrix1*} take the form
\begin{align}\label{eq:Pfaffian_qp_tildeK*}
    \tilde{K}_n = 
    \begin{pmatrix}
        K_n & & \\
        & -\mathbf{1}_{s_{\mathbf{l}_n}} & \\
        & & +\mathbf{1}_{s_{\mathbf{l}_n}}
    \end{pmatrix},
    \quad
    \tilde{\mathbf{t}}_n = 
    \begin{pmatrix}
        \mathbf{t} \\ 1 \\ \vdots \\ 1
    \end{pmatrix}.
\end{align}
Here, $\mathbf{1}_{s_{\mathbf{l}_n}}$ denotes the $s_{\mathbf{l}_n} \times s_{\mathbf{l}_n}$ identity matrix. We can then condense $\tilde{\mathbf{l}}_n=(\mathbf{l}_n^\mathsf{T},\underbrace{1,-1,\cdots,1,-1}_{2s_{\mathbf{l}_n}},\underbrace{0,\cdots,0}_{2s_{\mathbf{l}_n}})^\mathsf{T}$, whose topological spin and electric charge both vanish exactly. For example, for $n=3$, we have $\tilde{\mathbf{l}}_3=(\mathbf{l}_3^\mathsf{T},1,-1,1,-1,1,-1,0,\cdots,0)^\mathsf{T}$.

Now, we determine the $K$ matrix and the charge vector describing the theory obtained after condensation. The condensation of $\tilde{\mathbf{l}}_n$ is described by
\begin{align}
    \tilde{K}_{H,n} = 
    \begin{pmatrix}
        \tilde{K}_n & \tilde{\mathbf{l}}_n \\
        \tilde{\mathbf{l}}_n^\mathsf{T} & 0
    \end{pmatrix},
    \quad
    \tilde{\mathbf{t}}_{H,n} = 
    \begin{pmatrix}
        \tilde{\mathbf{t}}_n \\ 0
    \end{pmatrix}.
\end{align}
Any two $K$ matrices are equivalent under conjugation by an $\mathrm{SL}(N,\mathbb{Z})$ matrix \cite{wen1992classification}, where $N$ denotes the dimension of the $K$ matrix. To obtain the $K$ matrix describing the quasiparticle hierarchy states, we seek a transformation matrix $X_n \in \mathrm{SL}(N,\mathbb{Z})$ such that
\begin{align} \label{eq:transformed_K_t*}
    \tilde{K}_{qp, n} &\coloneqq X_n^\mathsf{T} \tilde{K}_{H,n} X_n =
    \begin{pmatrix}
        0 & 0 & 0 & 0\\
        0 & K_{qp,n} & 0 & 0 \\
        0 & 0 & -\mathbf{1}_{s_{\mathbf{l}_n}} & 0 \\
        0 & 0 & 0 & +\mathbf{1}_{s_{\mathbf{l}_n}}
    \end{pmatrix}, \nonumber \\
    \mathbf{t}_{qp,n} &\coloneqq X^\mathsf{T} \tilde{\mathbf{t}}_{H,n} = 
    \begin{pmatrix}
        0 & \mathbf{t}_{qp}^\mathsf{T} & 1 & \cdots & 1
    \end{pmatrix}^\mathsf{T}.
\end{align}
Here, $\pm \mathbf{1}_{s_{\mathbf{l}_n}\times s_{\mathbf{l}_n}}$ denote the trivial blocks introduced above, and the entries equal to $1$ associated with these blocks in the charge vector correspond to them. Upon discarding the trivial blocks, the $n$-th quasiparticle hierarchy state is described by $K_{qp, n}$ and $\mathbf{t}_{qp}$.

We claim that the following matrix $X_n \in \mathrm{SL}(N,\mathbb{Z})$ transforms $\tilde{K}_{H,n}$ into the desired form. Let the $i$-th column of $(X)_n$ be written as $(X_n)_i=(\mathbf{v}_i'^{\mathsf{T}} \ n_i)^\mathsf{T}$, where $\mathbf{v}_i'$ is an integer vector of length $N-1$ and $n_i \in \mathbb{Z}$. Then
\begin{equation} \label{eq:def_X*}
    \begin{split}
        \mathbf{v}_0' &= \text{ord} (\mathbf{l}_n) \tilde{K}_n^{-1} \tilde{\mathbf{l}}_n^\mathsf{T}, \quad n_0 =-\text{ord}(\tilde{\mathbf{l}}_n), \\
        \mathbf{v}_1' &= (\pm 1, 0 , \cdots, 0)^\mathsf{T}, \quad n_1=0,\\
        \mathbf{v}_2'&=(0,0,0,0,\underbrace{1,-1,...,1,-1}_{2s_{\mathbf{l}_n}},\underbrace{0,\cdots,0}_{2s_{\mathbf{l}_n}})^\mathsf{T}, \quad n_2=1, \\
        \mathbf{v}_i'&= \mathbf{e}_{i}, \quad n_i=0 \quad (3 \leq i \leq N)
    \end{split}    
\end{equation}
Here, $\mathbf{e}_{i}$ denotes the unit vector whose $i$-th entry is $1$ and all other entries are $0$. The sign of the first entry of $\mathbf{v}_1'$ is chosen such that $\det X_n = 1$.

To verify the claim, we use the fact, proved in the Appendix of Ref.~\cite{zhang2025hierarchy}, that the $(i,j)$-entry of $\tilde{K}_{qp, n}$, denoted by $(\tilde{K}_{qp,n})_{ij}$, can be interpreted as the mutual braiding phase between the anyons associated with the $i$-th and $j$-th columns of $X_n$. Define $\mathbf{v}_i = \tilde{K}_n \mathbf{v}_i'$ which labels a trivial anyon in the theory described by $\tilde{K}_n$. Then $(\tilde{K}_{qp,n})_{ij} = (\mathbf{v}_i + n_i \tilde{\mathbf{l}}_n)^\mathsf{T} \tilde{K}_n^{-1} (\mathbf{v}_j+n_j\tilde{\mathbf{l}}_n)$, which is precisely the mutual braiding phase between the anyons labeled by $\mathbf{c}_i\coloneqq(\mathbf{v}_i + n_i \tilde{\mathbf{l}}_n)^\mathsf{T}$ and $\mathbf{c}_j \coloneqq (\mathbf{v}_j + n_j \tilde{\mathbf{l}}_n)^\mathsf{T}$. Since the first row and the first column of $\tilde{K}_{qp, n}$ are required to vanish, we choose the trivial anyon $\mathbf{v}_0 = \text{ord}({\tilde{\mathbf{l}}_n}) \tilde{\mathbf{l}}_n$ so that $\mathbf{v}_0' = \text{ord}(\tilde{\mathbf{l}}_n) \tilde{K}_n^{-1} \tilde{\mathbf{l}}_n$ and $n_0 = - \text{ord}(\tilde{\mathbf{l}}_n)$. 

We first show that $\tilde{K}_{qp,n}$ has the desired form. Let $K_{qp, n}$ be an $M\times M$ matrix. Suppose $1 \leq i \leq M$ with $i \neq 2$, and $M + 1 \leq j \leq N$. By construction of $X_n$, the anyon corresponding to $\mathbf{c}_i$ contains only anyons in the theory described by $K_n$; that is, the $k$-th entry of column vector $\mathbf{c}_i$ vanishes for $k \geq M + 1$. Likewise, the anyon corresponding to $\mathbf{c}_j$ belongs entirely to one of the trivial sectors;that is, all entries of the column vector $\mathbf{c}_j$ vanish except for a single $k$-th entry with $M+1 \leq k \leq N$. Therefore, the mutual braiding phase between $\mathbf{c}_i$ and $\mathbf{c}_j$ vanishes, and hence  $(\tilde{K}_{qp, n})_{ij}=0$ for $1 \leq i \leq M$ with $i \neq 2$, and $M + 1 \leq j \leq N$. 

Now consider $i = 2$ and $M + 1 \leq j \leq N$. By construction, 
$\mathbf{v}_2 = \tilde{K}_n^{-1} \mathbf{v}_2'=(0, 0, 0, 0, \underbrace{-1, +1, \cdots, -1, +1}_{2 s_{\mathbf{l}_n}},\underbrace{0, \cdots, 0}_{2 s_{\mathbf{l}_n}})^\mathsf{T}$. It follows that $\mathbf{c}_2 = \mathbf{v}_2 + \tilde{\mathbf{l}}_n=(\mathbf{l}_n^\mathsf{T}, \underbrace{0, \cdots, 0}_{2 s_{\mathbf{l}_n}},\underbrace{0, \cdots, 0}_{2 s_{\mathbf{l}_n}},0)^\mathsf{T}$. Thus, $\mathbf{c}_2$ has no non-zero entries in the components corresponding to the $\nu = \pm 1$ blocks, and the braiding phase between the anyons labeled by $\mathbf{c}_2$ and $\mathbf{c}_j$ therefore vanishes. Hence, $(\tilde{K}_{qp, n})_{2 j}=0$ for $M + 1 \leq j \leq N$. 

Furthermore, $(\tilde{K}_{qp, n})_{ij}=\pm \delta_{ij}$ for $M + 1 \leq i, j \leq N$ because these entries correspond either to  braiding phase between nontrivial anyons belonging to distinct $\nu = \pm 1$ blocks, or to twice the topological spin of the nontrivial anyon in a single $\nu = \pm 1$ blocks, which is $\pm \frac{1}{2}$. 

Finally, we show that the transformed charge vector $X_n^\mathsf{T} \tilde{\mathbf{t}}_{H,n}$ takes the form given in Eq.~\eqref{eq:transformed_K_t*}. Its $i$-th entry is
\begin{equation}
    (X_n^{\mathsf{T}} \tilde{\mathbf{t}}_{H,n})_i = \mathbf{v}_i'^{\mathsf{T}} \tilde{\mathbf{t}}_n.
\end{equation}
Hence, for $M + 1 \leq i \leq N$, $(X_n^\mathsf{T} \tilde{\mathbf{t}}_{H,n})_i = 1$ by construction of $\tilde{\mathbf{t}}_n$ and $\mathbf{v}_i'$. For $i=0$, we find
\begin{equation}
    (X_n^\mathsf{T} \tilde{\mathbf{t}}_{H,n})_0 = \mathbf{v}_0^\mathsf{T} \tilde{K}_n^{-1} \tilde{\mathbf{t}}_n = q_{\mathbf{v}_0},
\end{equation}
which is the electric charge of the anyon labeled by $\mathbf{v}_0=\text{ord}(\tilde{\mathbf{l}}_n)\tilde{\mathbf{l}}_n$. Therefore, $(X_n^\mathsf{T} \tilde{\mathbf{t}}_{H,n})_0=0$.

Thus, with the transformation matrix $X_n$ defined in ~\eqref{eq:def_X*}, $\tilde{K}_{qp,n}$ indeed takes the desired form. Moreover, each entry of $K_{qp,n}$ is obtained by computing the braiding phase between $\mathbf{c}_i$ and $\mathbf{c}_j$ for $1\leq i, j \leq M$. The charge vector $\mathbf{t}_{qp}$ is obtained from $X_n^\mathsf{T} \tilde{\mathbf{t}}_{H,n}$. We therefore arrive at the following $K$ matrix and charge vector for the $n$-th quasiparticle hierarchy state of the Pfaffian state:
\begin{equation}
\begin{split}
    K_{qp,n} &= 
    \begin{pmatrix}
        -1 & 0 & 1 & 0 \\
        0 & 2n & -2n+1 & -1 \\
        1 & -2n+1 & 2n & -1 \\
        0 & -1 & -1 & 4
    \end{pmatrix},
    \\
    \mathbf{t}_{qp} &=
    \begin{pmatrix}
        -1 & 0 & 0 & 0
    \end{pmatrix}^\mathsf{T}.    
\end{split}
\end{equation}
The total quantum dimension, filling fraction, and chiral central charge are given by
\begin{equation}
    D_n=\sqrt{2(16n-3)}, \quad \nu_n=\frac{8n-1}{16n-3},\quad c_{-}=3,
\end{equation}
as expected.

We finally note that the procedure used to obtain the quasiparticle hierarchy states of the Pfaffian state can be extended to other Pfaffian-type states, such as the anti-Pfaffian and bosonic Pfaffian state. A slight modification is required when the topological spin of the condensate is negative, $s_{\mathbf{l}_n}<0$. In this case, one simply interchanges the $-\mathbf{1}_{s_{\mathbf{l}_n}}$ and $+\mathbf{1}_{s_{\mathbf{l}_n}}$ in $\tilde{K}_n$, while all other quantities are defined in the same way, so that $s_{\tilde{\mathbf{l}}_n}=0$. Here, $\mathbf{1}_{s_{\mathbf{l}_n}}$ is understood to denote the $|s_{\mathbf{l}_n}|\times |s_{\mathbf{l}_n}|$ identity matrix.

\section{Other hierarchies of the Pfaffian state}
In the main text, we derived one quasihole hierarchy and one quasiparticle hierarchy of the Pfaffian state. In this section, we consider other quasihole and quasiparticle hierarchies of the Pfaffian state suggested in Ref.~\cite{zhang2025hierarchy} whose filling fractions are different from those discussed in the main text. We will show that the filling fraction, chiral central charge, total quantum dimension, and the anyon content of each hierarchy state agree with Ref.~\cite{zhang2025hierarchy}.

We first construct other quasihole hierarchy of the Pfaffian state. The Lagrangian describing the quasihole hierarchy of the Pfaffian state discussed in the main text is
\begin{align} \label{eq:pfaffian_quasihole*}
    \mathcal{L} &= \frac{1}{4 \pi} (a_\uparrow d a_\uparrow + a_\downarrow d a_\downarrow) + \frac{3}{2 \pi} a_\uparrow d a_\downarrow + \frac{1}{2 \pi} A d (a_\uparrow + a_\downarrow) \nonumber \\
    &\quad  + \mathrm{CS}[A, g] + \frac{1}{2 \pi} a_\uparrow d \alpha + \frac{2 n}{4 \pi} \alpha d \alpha.
\end{align}
We now reverse the sign of the Chern-Simons term for $\alpha$ in Eq.~\eqref{eq:pfaffian_quasihole*}, namely $n \to - n$. We then introduce an additional current $\star j'=\frac{1}{2\pi}d\beta$ that carries charge-1 under $\alpha$, together with a level-4 Chern-Simons term for $\beta$. As before, $\beta$ is an emergent $\mathrm{U(1)}$ gauge field. In the $K$-matrix formalism, the resulting theory is described by
\begin{equation}
    K_n' = 
    \begin{pmatrix}
        -1 & -3 & 1 & 0 \\
        -3 & -1 & 0 & 0 \\
        1 & 0 & -2n & 1 \\
        0 & 0 & 1 & -4
    \end{pmatrix},
    \quad
    \mathbf{t}' = 
    \begin{pmatrix}
        -1 \\ -1 \\ 0 \\ 0
    \end{pmatrix}.
\end{equation}
We then condense a charge-neutral order-2 boson labeled by $\mathbf{l}_n = (0, 1, 2 n, -1)^\mathsf{T}$. After condensation, the theory is described by
\begin{equation}
    K_{qh, n} = 
    \begin{pmatrix}
        -1 & 0 & 1 & 0 \\
        0 & -2n & 2n & -1 \\
        1 & 2n & -2n & -1 \\
        0 & -1 & -1 & -4
    \end{pmatrix},
    \quad
    \mathbf{t}_{qh} =
    \begin{pmatrix}
        1 \\ 0 \\ 0 \\ 0 
    \end{pmatrix}.
\end{equation}
The corresponding filling fraction, chiral central charge, and total quantum dimension are
\begin{equation}
    \nu_n = \frac{8 n - 1}{16 n - 1}, \quad c_{-} = - 1, \quad D_n = \sqrt{2(16 n - 1)},
\end{equation}
respectively. Moreover, the anyon content is generated by the Abelian anyon $(0, 0, 4 n - 1, 8 n - 3)^\mathsf{T}$. Its topological spin is $\frac{16 n - 3}{2 (16 n - 1)}$, and its electric charge is $\frac{1}{16 n - 1}$ in units of the electron charge. (We will henceforth set $e = 1$ for notational convenience.)

Other quasiparticle state involves a higher-order hierarchy construction. Specifically, the $K$ matrix and the charge vector are given by
\begin{equation}
    K_n =
    \begin{pmatrix}
    -1 & -3 & 1 & 0 & 0 \\
    -3 & -1 & 0 & 0 & 0 \\
    1 & 0 & 2n & 1 & 1 \\
    0 & 0 & 1 & 4 & 0 \\
    0 & 0 & 1 & 0 & 4 \\
    \end{pmatrix},
    \quad
    \mathbf{t}= 
    \begin{pmatrix}
        -1 \\ -1 \\ 0 \\ 0 \\ 0
    \end{pmatrix},
\end{equation}
In this case, we condense two distinct charge-neutral bosons of order-2: $\mathbf{l}_n^{(1)}=(0, 1, 2 n - 1, 1, -1)^\mathsf{T}$ and $\mathbf{l}_n^{(2)}=(0, 1, 2 n - 1, - 1, 1)^\mathsf{T}$. Both have exactly zero electric charge, but their topological spins are $s_{\mathbf{l}_n^{(1)}} = s_{\mathbf{l}_n^{(2)}} = n$. Since in the previous method, we condense only a single boson, it must be generalized to accommodate the condensation of both bosons.

As before, we stack an appropriate number of trivial blocks onto the $K$ matrix to obtain $\tilde{K}_n$. We then adjoin $\tilde{\mathbf{l}}_1^{(1)}$ and $\tilde{\mathbf{l}}_1^{(2)}$ to $\tilde{K}_n$, yielding
\begin{equation}
    \tilde{K}_{H,n}=
    \begin{pmatrix}
        \tilde{K}_n & \tilde{\mathbf{l}}_n^{(1)} & \tilde{\mathbf{l}}_n^{(2)} \\
        \tilde{\mathbf{l}}_{n}^{(1)\mathsf{T}} & 0 & 0 \\
        \tilde{\mathbf{l}}_n^{(2)\mathsf{T}} & 0 & 0
    \end{pmatrix}.
\end{equation}
Define $L= \begin{pmatrix} \tilde{\mathbf{l}}_n^{(1)} & \tilde{\mathbf{l}}_n^{(2)} \end{pmatrix}.$ The determinant of $\tilde{K}_{H,n}$ is then $\det \tilde{K}_{H,n} =(\det\tilde{K}_n) \left(\det \left[-L^\mathsf{T} \tilde{K}^{-1}_n L\right]\right)$. The diagonal entries of $L^\mathsf{T} \tilde{K}_n^{-1} L$ are equal to $2 s_{\mathbf{l}_n^{(1)}}$ and $2 s_{\mathbf{l}_n^{(2)}}$, respectively. Moreover, the off-diagonal entries are given by the mutual braiding phase between $\mathbf{l}_n^{(1)}$ and $\mathbf{l}_n^{(2)}$. Since $\det\tilde{K}_n \neq 0$, we require $\det \left[-L^\mathsf{T} \tilde{K}_n^{-1} L\right] = 0$. Thus, we must arrange for not only the topological spins of the two bosons, but also their mutual braiding phase, to vanish exactly. The mutual braiding phase between $\tilde{\mathbf{l}}_n^{(1)}$ and $\tilde{\mathbf{l}}_n^{(2)}$ is $2\pi \times (2 n - 1)$. We therefore stack $2 n + 1$ copies of the $\nu = - 1$ block and $2 n + 1$ copies of $\nu = +1$ block, and condense
\begin{equation}
    \begin{split}
        \tilde{\mathbf{l}}_n^{(1)} &=(\mathbf{l}_n^{(1)\mathsf{T}}, \underbrace{1,-1,\cdots ,1,-1}_{2n-2},1,-1,0,\underbrace{0,\cdots ,0}_{2n+1})^\mathsf{T},\\
        \tilde{\mathbf{l}}_n^{(2)} &= (\mathbf{l}_n^{(2)\mathsf{T}}, \underbrace{1,-1,\cdots ,1,-1}_{2n-2},1,0,-1,\underbrace{0,\cdots ,0}_{2n+1})^\mathsf{T}
    \end{split}
\end{equation}
Here, both $\tilde{\mathbf{l}}_n^{(1)}$ and $\tilde{\mathbf{l}}_n^{(2)}$ have exactly zero topological spins and zero electric charges, and their mutual braiding phase also vanishes exactly. 

Let the $i$-th column of the transformation matrix $X_n \in \mathrm{SL}(N, \mathbb{Z})$ be written as ${(X_{n})}_i=(\mathbf{v}_i'^{\mathsf{T}} \ n_i \ m_i)^\mathsf{T}$, where $\mathbf{v}_i'=\tilde{K}^{-1} \mathbf{v}_i$ and $\mathbf{v}_i$ labels a trivial anyon in the theory described by $\tilde{K}_n$. By a calculation similar to the one above, the $(i,j)$-entry of $\tilde{K}_{qp,n}$ is then given by the mutual braiding phase between the anyons $\mathbf{c}_i \coloneqq \mathbf{v}_i + n_i \tilde{\mathbf{l}}_n^{(1)} + m_i \tilde{\mathbf{l}}_n^{(2)}$ and $\mathbf{c}_j \coloneqq \mathbf{v}_j + n_j \tilde{\mathbf{l}}_n^{(1)} + m_j \tilde{\mathbf{l}}_n^{(2)}$. Equivalently, $(\tilde{K}_{qp,n})_{ij} = \mathbf{c}_i^\mathsf{T} \tilde{K}_n^{-1} \mathbf{c}_j$. Here, we have used the fact that $\tilde{\mathbf{l}}_n^{(i) \mathsf{T}} \tilde{K}_n^{-1} \tilde{\mathbf{l}}_n^{(j)} = 0$ for all $i$ and $j$. Since we require the first two rows and columns of $\tilde{K}_{qp,n}$ to vanish, $\mathbf{c}_0$ and $\mathbf{c}_1$ must represent the vacuum. We therefore choose $\mathbf{v}_0 = \text{ord}(\tilde{\mathbf{l}}_n^{(1)}) \tilde{\mathbf{l}}_n^{(1)} \rightarrow \mathbf{v}_0' = \text{ord}(\tilde{\mathbf{l}}_n^{(1)}) \tilde{K}_n^{-1} \tilde{\mathbf{l}}_n^{(1)}$, with $n_0 = -\text{ord}(\tilde{\mathbf{l}}_n^{(1)})$ and $m_0 = 0$. Similarly, we choose $\mathbf{v}_1 = \text{ord}(\tilde{\mathbf{l}}_n^{(2)}) \tilde{\mathbf{l}}_n^{(2)} \rightarrow \mathbf{v}_1' = \text{ord}(\tilde{\mathbf{l}}_n^{(2)}) \tilde{K}^{-1} \tilde{\mathbf{l}}_n^{(2)}$, with $n_2 = 0$ and $m_2 = - \text{ord}(\tilde{\mathbf{l}}_n^{(2)})$. Note that a similar calculation appeared in Ref.~\cite{zhang2025hierarchy}. We have extended the discussion to the case of condensing two bosons.

We choose the remaining columns as follows:
\begin{equation} \label{eq:def_X_otherqp*}
    \begin{split}
        \mathbf{v}_2' &= (0,\cdots,0,\underbrace{1,-1, \cdots,1,-1}_{2n-2},1,-1,0,\underbrace{0,\cdots,0}_{2n+1})^\mathsf{T} \\
        n_2 &= 1, \quad m_2 = 0, \\
        \mathbf{v}_3' &= (0,\cdots,0,\underbrace{1,-1, \cdots,1,-1}_{2n-2},1,0,-1,\underbrace{0,\cdots,0}_{2n+1})^\mathsf{T}, \\
        n_2 &= 0, \quad m_2 = 1\\
        \mathbf{v}_i'&=\mathbf{e}_{i-4},\quad n_i =m_i=0, \quad (4\leq i \leq 6) \\
        \mathbf{v}_i'&=\mathbf{e}_{i-2},\quad n_i =m_i=0, \quad (7\leq i \leq N)
    \end{split}
\end{equation}
Then, under the $\mathrm{SL}(N,\mathbb{Z})$ transformation generated by $X_n$, we obtain
\begin{equation}
\begin{split}
    \tilde{K}_{qp,n} &\coloneqq X_n^\mathsf{T} \tilde{K}_{H,n} X_n \\
    &= 
    \begin{pmatrix}
        0 & 0 & 0 & 0 & 0 \\
        0 & 0 & 0 & 0 & 0 \\
        0 & 0 & K_{qp,n} & 0 & 0 \\
        0 & 0 & 0 & -\mathbf{1}_{2 n + 1} & 0 \\
        0 & 0 & 0 & 0 & +\mathbf{1}_{2 n + 1}
    \end{pmatrix},
\end{split}
\end{equation}
as desired. The proof proceeds in the same way as in the previous case. 

As a result, we obtain
\begin{equation}
\begin{split}
    K_{qp,n} &= 
    \begin{pmatrix}
        2n & 2n-1 & 0 & 1 & 2n-1 \\
        2n-1 & 2n & 0 & 1 & 2n-1 \\
        0 & 0 & -1 & -3 & 1 \\
        1 & 1 & -3 & -1 & 0 \\
        2n-1 & 2n-1 & 1 & 0 & 2n
    \end{pmatrix}, \\
    \mathbf{t}_{qp} &= 
    \begin{pmatrix}
        0 & 0 & -1 & -1 & 0
    \end{pmatrix}^\mathsf{T}
\end{split}
\end{equation}
The corresponding filling fraction, chiral central charge, and total quantum dimension are
\begin{equation}
    \nu_n = \frac{8 n - 2}{16 n - 5}, \quad c_{-} = 4, \quad D_n=\sqrt{2 (16 n - 5)}.
\end{equation}
The anyon content is generated by the Abelian anyon $(0, 0, 4 n - 2, - 4 n + 3, 0)^\mathsf{T}$. Its topological spin is $\frac{16 n - 3}{2 (16 n - 5)}$, and its electric charge is $\frac{1}{16 n - 5}$.  

\section{Other Pfaffian-type states}
In this section, we construct quasihole and quasiparticle hierarchies of the anti-Pfaffian, bosonic Pfaffian, and PH-Pfaffian states. Again, the resulting values of $\nu_n$, $c_{-}$, and $D_n$ agree with those in Ref.~\cite{zhang2025hierarchy}.

\subsection{Anti-Pfaffian state}
The anti-Pfaffian state \cite{levin2007particle-hole,lee2007particle-hole} is described by the Lagrangian
\begin{equation}
    \begin{split}
        \mathcal{L} &= \frac{2}{4 \pi} \Tr\!\left[a d a + \frac{2}{3} a^3\right] - \frac{3}{4 \pi} (\Tr a) d (\Tr a) \\ 
        & \quad + \frac{1}{2 \pi} (\Tr a) d A,
    \end{split}
\end{equation}
where $a$ is a dynamical $\mathrm{U(2)}$ gauge field and $A$ is the background electromagnetic field. If the $\mathrm{U(2)}$ gauge group is broken down to $\mathrm{U(1)} \times \mathrm{U(1)}$, then $a$ is restricted to $a \to \diag(a_\uparrow, a_\downarrow)$. The Lagrangian then becomes
\begin{equation}
    \begin{split}
        \mathcal{L} &\to -\frac{1}{4\pi}(a_\uparrow da_\uparrow + a_\downarrow da_\downarrow) -\frac{3}{2\pi} a_\uparrow da_\downarrow \\
        & \quad + \frac{1}{2\pi} (a_\uparrow + a_\downarrow) dA.
    \end{split}
\end{equation}
In the $K$-matrix formalism, this Lagrangian is represented by
\begin{equation}
    K = 
    \begin{pmatrix}
        1 & 3 \\
        3 & 1
    \end{pmatrix},
    \quad
    \mathbf{t} =
    \begin{pmatrix}
        1 \\ 1
    \end{pmatrix}.
\end{equation}
We first consider the quasihole hierarchy of the anti-Pfaffian state. Performing the hierarchy construction as before, we obtain the following $K$-matrix and charge vector:
\begin{equation}
    K_{qh, n} = 
    \begin{pmatrix}
        1 & 3 & -1 \\
        3 & 1 & 0 \\
        -1 & 0 & 2n
    \end{pmatrix},
    \quad
    \mathbf{t} =
    \begin{pmatrix}
        1 \\ 1 \\ 0
    \end{pmatrix}.
\end{equation}
The corresponding filling fraction, chiral central charge, and total quantum dimension are
\begin{equation}
    \nu_n = \frac{8 n + 1}{16 n + 1}, \quad c_{-} = 1, \quad D_n = \sqrt{2 (16 n + 1)}.
\end{equation}
Moreover, the anyon content is generated by the Abelian anyon labeled by $(0, 1, -2 n)^\mathsf{T}$. Its topological spin is $\frac{16 n + 3}{2 (16 n + 1)}$, and its electric charge is $\frac{1}{16 n + 1}$.

The quasiparticle hierarchy is obtained through a second order hierarchy construction, with $K$ matrix and charge vector
\begin{equation}
    K_n=
    \begin{pmatrix}
         1 & 3 & -1 & 0 \\ 
         3 & 1 & 0 & 0 \\ 
         -1 & 0 & -2n & -1 \\
         0 & 0 & -1 & -4
     \end{pmatrix},
     \quad
     \mathbf{t}=
     \begin{pmatrix}
         1 \\ 1 \\ 0 \\ 0
     \end{pmatrix}.
\end{equation}
In this case, we condense the boson $\mathbf{l}_n=(0,1,2n-1,-1)^\mathsf{T}$ whose electric charge is exactly zero. Its topological spin is $s_{\mathbf{l}_n} = -n$, which is not exactly 0. We therefore stack $2 n$ copies of the $\nu = +1$ block and $2 n$ copies of the $\nu = -1$ block.

The resulting theory is described by
\begin{equation}
\begin{split}
    K_{qp,n} &=
    \begin{pmatrix}
        1 & 0 & -1 & 0 \\
        0 & -2n & 2n-1 & -1 \\
        -1 & 2n-1 & -2n & -1 \\
        0 & -1 & -1 & -4
    \end{pmatrix},
    \\
    \mathbf{t}_{qp} &=
    \begin{pmatrix}
        1 & 0 & 0 & 0
    \end{pmatrix}^\mathsf{T}.    
\end{split}
\end{equation}
The corresponding filling fraction, chiral central charge, and total quantum dimension are
\begin{equation}
    \nu_n = \frac{8 n - 2}{16 n - 3},\quad c_{-}=-2, \quad D_n = \sqrt{2 (16 n - 3)}.
\end{equation}
The anyon content is generated by the Abelian anyon labeled by $(0, 0, 4 n - 2, 8 n - 3)^\mathsf{T}$. Its topological spin is $\frac{16 n - 5}{2 (16 n - 3)}$, and its electric charge is $\frac{1}{16 n -3}$.

\subsection{Bosonic Pfaffian state}
The bosonic Pfaffian state \cite{fradkin1998chern-simons} is described by the effective Lagrangian
\begin{equation}
    \begin{split}
        \mathcal{L} &= -\frac{2}{4 \pi} \Tr\!\left[a d a + \frac{2}{3} a^3 \right] + \frac{2}{4 \pi} (\Tr a) d (\Tr a) \\
        & \quad + \frac{1}{2 \pi} (\Tr a) d b + \frac{1}{2 \pi} A d b,
    \end{split}
\end{equation}
where $a$ and $b$ are dynamical $\mathrm{U(2)}$ and $\mathrm{U(1)}$ gauge fields, respectively, and $A$ is the background electromagnetic field. If the $\mathrm{U}(2)$ gauge group breaks down to $\mathrm{U(1)} \times \mathrm{U(1)}$, $a$ is restricted to $a \to \diag(a_\uparrow, a_\downarrow)$. The Lagrangian then becomes
\begin{equation}
    \begin{split}
        \mathcal{L} \to \frac{2}{2\pi} a_\uparrow  da_\downarrow + \frac{1}{2\pi}(a_\uparrow + a_\downarrow ) db + \frac{1}{2\pi} A db.
    \end{split}
\end{equation}
In the $K$-matrix formalism, this theory is described by
\begin{equation}
    K = 
    \begin{pmatrix}
        0 & -2 & -1 \\
        -2 & 0 & -1 \\
        -1 & -1 & 0
    \end{pmatrix},
    \quad
    \mathbf{t} = 
    \begin{pmatrix}
        0 \\ 0 \\ 1
    \end{pmatrix}.
\end{equation}

The quasihole hierarchy is obtained by performing the hierarchy construction once, yielding
\begin{equation}
    K_{qh, n} = 
    \begin{pmatrix}
        0 & -2 & -1 & -1 \\ 
        -2 & 0 & -1 & 0 \\
        -1 & -1 & 0 & 0 \\
        -1 & 0 & 0 & -2n 
    \end{pmatrix},
    \quad
    \mathbf{t} = 
    \begin{pmatrix}
        0 \\ 0 \\ 1 \\ 0
    \end{pmatrix},
\end{equation}
where $n \in \mathbb{N}$. The corresponding filling fraction, chiral central charge, and total quantum dimension are
\begin{equation}
    \nu_n = \frac{8 n}{8 n + 1}, \quad c_{-} = 0, \quad D_n = \sqrt{8 n + 1}.
\end{equation}
The anyon content is generated by the Abelian anyon $(-1, 1, 0, 1)^\mathsf{T}$. Its topological spin is $\frac{4n}{8 n + 1}$, and its electric charge is $\frac{1}{8 n + 1}$.

The quasiparticle hierarchy is obtained through a second-order hierarchy construction:
\begin{equation}
    K_{n} =\begin{pmatrix}
         0 & -2 & -1 & -1 & 0 \\ -2 & 0 & -1 & 0 & 0 \\ -1 & -1 & 0 & 0 & 0 \\ -1 & 0 & 0 & 2n & -1 \\ 0 & 0 & 0 & -1 & 4
     \end{pmatrix},\quad 
     \mathbf{t} = \begin{pmatrix}
         0 \\ 0 \\ 1 \\ 0 \\ 0
     \end{pmatrix}.
\end{equation}
We condense $\mathbf{l}_n=(0,1,1,2n,1)^\mathsf{T}$ whose electric charge is exactly 0. Since its topological spin is $s_{\mathbf{l}_n} = n$, we stack $2 n$ copies of the $\nu = -1$ block and $2 n$ copies of the $\nu = +1$ block. After condensation, the resulting theory is described by
\begin{equation}
    K_{qp,n} =
    \begin{pmatrix}
        0 & 0 & 1 & 1 & 0 \\
        0 & 2n & 1 & 2n & 1 \\
        1 & 1 & 0 & 0 & 0 \\
        1 & 2n & 0 & 2n & -1 \\
        0 & 1 & 0 & -1 & 4
    \end{pmatrix},
    \quad
    \mathbf{t}_{qp} = 
    \begin{pmatrix}
        0 \\ 0 \\ 1 \\ 0 \\ 0
    \end{pmatrix}.
\end{equation}
The corresponding total quantum dimension, the filling fraction, and the chiral central charge are
\begin{equation}
    D_n = \sqrt{8 n - 2}, \quad \nu_n = \frac{8 n - 1}{8 n - 2},\quad c_{-} = 3.
\end{equation}
The anyon content is generated by the Abelian anyon $(0, 0, 1, 0, 2)^\mathsf{T}$. Its topological spin is $\frac{8 n - 1}{2 (8 n - 2)}$, and its electric charge is $\frac{1}{8 n - 2}$.

\subsection{PH-Pfaffian state}
The PH-Pfaffian \cite{son2015is} state is described by the effective Lagrangian
\begin{equation}
    \begin{split}
        \mathcal{L} &= \frac{2}{4 \pi} \Tr\!\left[ a d a + \frac{2}{3} a^3 \right] - \frac{2}{4 \pi} (\Tr a) d (\Tr a) \\ 
        & \quad + \frac{1}{2\pi} (\Tr a) db - \frac{3}{4 \pi} b d b - \frac{1}{2 \pi} A d b, 
    \end{split}
\end{equation}
where $a$, $b$, and $A$ are defined as before. Upon breaking the $\mathrm{U(2)}$ gauge group down to $\mathrm{U(1)} \times \mathrm{U(1)}$, the Lagrangian becomes
\begin{equation}
    \begin{split}
        \mathcal{L} \to -\frac{2}{2 \pi} a_\uparrow d a_\downarrow + \frac{1}{2 \pi} (a_\uparrow + a_\downarrow) d b - \frac{3}{4 \pi} b d b - \frac{1}{2 \pi} A d b.
    \end{split}
\end{equation}
In the $K$-matrix formalism, this theory is described by
\begin{equation}
    K = 
    \begin{pmatrix}
        0 & 2 & -1  \\
        2 & 0 & -1 \\
        -1 & -1 & 3 \\
    \end{pmatrix},
    \quad
    \mathbf{t} = 
    \begin{pmatrix}
        0 \\ 0 \\ -1
    \end{pmatrix}.
\end{equation}

The quasihole hierarchy is obtained by performing the hierarchy construction once:
\begin{equation}\label{eq:ph_pfaffian_qh1*}
    K_{qh, n} = 
    \begin{pmatrix}
        0 & 2 & -1 & -1 \\
        2 & 0 & -1 & 0 \\
        -1 & -1 & 3 & 0 \\
        -1 & 0 & 0 & -2n
    \end{pmatrix},
    \quad
    \mathbf{t}_{qh} = 
    \begin{pmatrix}
        0 \\ 0 \\ -1 \\ 0
    \end{pmatrix}.
\end{equation}
The corresponding filling fraction, chiral central charge, and total quantum dimension are
\begin{equation}
    \nu_n = \frac{8 n}{16 n + 1}, \quad c_{-}=0, \quad D_n=\sqrt{2(16n+1)}.
\end{equation}
The anyon content is generated by the Abelian anyon $(-1, 1, 0, 0)^\mathsf{T}$. Its topological spin is $\frac{16 n - 1}{2 (16 n + 1)}$, and its electric charge is $\frac{1}{16 n + 1}$.

Furthermore, other quasihole hierarchy of the PH-Pfaffian state, suggested in Ref.~\cite{zhang2025hierarchy}, can be obtained by replacing $n \to -n$ and performing the hierarchy construction once more. Specifically, the $K$-matrix and charge vector are given by
\begin{equation}\label{eq:ph_pfaffian_qh2*}
    K_n' = 
    \begin{pmatrix}
        0 & 2 & -1 & -1 & 0 \\
        2 & 0 & -1 & 0 & 0 \\
        -1 & -1 & 3 & 0 & 0 \\
        -1 & 0 & 0 & -2n & -1 \\
        0 & 0 & 0 & -1 & -4
    \end{pmatrix}, 
    \quad
    \mathbf{t}' = 
    \begin{pmatrix}
        0 \\ 0 \\ -1 \\ 0 \\ 0
    \end{pmatrix}.    
\end{equation}
We then condense the anyon labeled by $\mathbf{l}_n = (0, 1, -1, -2n, 1)^\mathsf{T}$. Its electric charge is exactly zero, and its topological spin is $s_{\mathbf{l}_n} = - n$. Thus, we stack $n$ copies of the $\nu=+1$ block and $n$ copies of the $\nu = - 1$ block. After condensation, the resulting theory is described by
\begin{equation}
\begin{split}
    K_{qh, n}' &= 
    \begin{pmatrix}
        0 & 0 & -1 & -1 & 0 \\
        0 & -2n & -1 & -2n & 1 \\
        -1 & -1 & 3 & 0 & 0 \\
        -1 & -2n & 0 & -2n & -1 \\
        0 & 1 & 0 & -1 & -4
    \end{pmatrix},
    \\
    \mathbf{t}_{qh}' &= 
    \begin{pmatrix}
        0 & 0 & -1 & 0 & 0
    \end{pmatrix}^\mathsf{T}.  
\end{split}
\end{equation}
The corresponding filling fraction, chiral central charge, and the total quantum dimension are
\begin{equation}
    \nu_n = \frac{8 n - 1}{16 n - 1}, \quad c_{-} = - 1, \quad D_n = \sqrt{2 (16 n - 1)}.
\end{equation} 
The anyon content is generated by the Abelian anyon $(0,0,0,-1,-2)^\mathsf{T}$ whose topological spin and charge are $\frac{16 n - 3}{2 (16 n - 1)}$ and $- \frac{1}{16 n - 1}$, respectively. 

We now consider the quasiparticle hierarchy of the PH-Pfaffian state. One such hierarchy state is obtained by replacing $n \to -n$ in Eq.~\eqref{eq:ph_pfaffian_qh1*}, while keeping the charge vector unchanged. The corresponding filling fraction, chiral central charge, and total quantum dimension are
\begin{equation}
    \nu_n = \frac{8 n}{16 n - 1}, \quad c_{-}=2, \quad D_n = \sqrt{2 ( 16 n - 1)}.
\end{equation}
The anyon content is generated by the Abelian anyon $(0, 1, 0, -2 n - 1)^\mathsf{T}$. Its topological spin and charge are $\frac{16 n + 1}{2 ( 16 n - 1)}$ and $\frac{1}{16 n - 1}$, respectively.

We now construct other quasiparticle hierarchy suggested in Ref.~\cite{zhang2025hierarchy}. This hierarchy is also obtained by replacing $n \to -n$ in Eq.~\eqref{eq:ph_pfaffian_qh2*} and then condensing the anyon $\mathbf{l}_n = (0, 1, -1, 2 n - 1, -3)^\mathsf{T}$. Its electric charge is exactly 0, while its topological spin is $s_{\mathbf{l}_n} = n - 1$. We therefore stack $2 (n - 1)$ copies of the $\nu = - 1$ block and $2 (n - 1)$ copies of the $\nu = + 1$ block. After condensation, the resulting theory is described by
\begin{equation}
\begin{split}
    K_{qp,n}' &=
    \begin{pmatrix}
        0 & 0 & -1 & -1 & 0 \\
        0 & 2n-2 & -1 & 2n-1 & -3 \\
        -1 & -1 & 3 & 0 & 0 \\
        -1 & 2n-1 & 0 & 2n & -1 \\
        0 & -3 & 0 & -1 & -4
    \end{pmatrix},
    \\
    \mathbf{t}_{qp}' &= 
    \begin{pmatrix}
        0 & 0 & 1 & 0 & 0
    \end{pmatrix}^\mathsf{T}.    
\end{split}
\end{equation}
The corresponding filling fraction, chiral central charge, and total quantum dimension are
\begin{equation}
    \nu_n = \frac{8 n + 1}{16 n + 1},\quad c_{-} = 1, \quad D_n = \sqrt{2 (16 n + 1)}.
\end{equation}
The anyon content is generated by the Abelian anyon $(1,0,0,0,0)^\mathsf{T}$. Its topological spin is $-\frac{16 n - 1}{2 (16 n + 1)}$, and its electric charge is $- \frac{1}{16 n + 1}$.

\section{Non-Abelian hierarchies of \texorpdfstring{$\nu = 1 / 3$}{nu = 1 / 3} Laughlin state}

In the main text, we have constructed a Chern-Simons-Ginzburg-Landau theory of the hierarchical anti-Read-Rezayi sequence emerging from the $\nu = 2 / 3$ Jain state. The Chern-Simons-Ginzburg-Landau theories for the non-Abelian hierarchical sequence starting from the $\nu = 1 / 3$ Laughlin state can also be described similarly. 

In Ref.~\cite{yutushui2025non-abelian}, the consecutive condensation of the Laughlin quasiparticles of the $\nu = 1 / 3$ Laughlin state yields the $\nu = 2 / 5$ Jain state and then the $\mathrm{SU}(2)_k$ topological orders for integer $k \geq 2$. The $\nu = 1 / 3$ Laughlin state can be described by the $\mathrm{U}(2)_{0, 12}$ Chern-Simons-Ginzburg-Landau theory
\begin{align}
   \mathcal{L} = - \frac{3}{4 \pi} (\Tr a) d (\Tr a) + \frac{1}{2 \pi} A d (\Tr a) + \mathcal{L}[\Phi, a].
\end{align}
Similarly as before, we add a $\mathrm{U}(2)$ Chern-Simons term $- \frac{1}{4 \pi} \Tr\!\left[b d b + \frac{2}{3} b^3\right] + \frac{1}{4 \pi} (\Tr b) d (\Tr b)$, where $b$ is a dynamical $\mathrm{U}(2)$ gauge field and $\Phi$ becomes the bifundamental representation of $a$ and $b$. After the condensation of $\Phi$, we get
\begin{align}
   \mathcal{L} &= - \frac{1}{4 \pi} \Tr\!\left[a d a + \frac{2}{3} a^3\right] - \frac{2}{4 \pi} (\Tr a) d (\Tr a) \nonumber \\
   &\quad + \frac{1}{2 \pi} A d (\Tr a),
\end{align}
which describes the $\nu = 2 / 5$ Jain state. By repeating the procedure, we finally obtain 
\begin{align}
   \mathcal{L} &= - \frac{k}{4 \pi} \Tr\!\left[a d a + \frac{2}{3} a^3\right] + \frac{k - 3}{4 \pi} (\Tr a) d (\Tr a) \nonumber \\
   &\quad + \frac{1}{2 \pi} A d (\Tr a).
\end{align}
For $k \geq 2$, the Lagrangian describes the $\mathrm{SU}(2)_k$ theory \cite{blok1992many} with filling fraction $\nu = \frac{2}{6 - 2k}$. This aligns with the wavefunction analysis of Ref.~\cite{yutushui2025non-abelian}.
\end{document}